\documentstyle[aps,prl,preprint,tighten,psfig]{revtex}

\newcommand{\beq}{\begin{equation}}
\newcommand{\eeq}[1]{\label{#1} \end{equation}}

\begin{document}
\draft
\preprint{CU-TP-880}

\title{ Observation of Partial $U_A(1)$ Restoration from Two-Pion
Bose-Einstein Correlations }

\author{ S.E. Vance$^1$, T. Cs\"{o}rg\H{o}$^{1,2}$ and D. Kharzeev$^3$}   

\address{$^1$Physics Department, Columbia University, New York, NY 10027\\
$^2$MTA KFKI RMKI, H-1525 Budapest 144, P.O. Box 49, Hungary\\
$^3$RIKEN-BNL Research Center, Brookhaven National Laboratory, 
Upton, NY 11973}
\date{24th February 1998}
\maketitle 

\begin{abstract}
The effective intercept parameter of 
the two-pion Bose-Einstein correlation function, $\lambda_*$, is
found to be sensitive to the partial restoration of $U_A(1)$ symmetry in
ultra-relativistic nuclear collisions.  An increase 
in the yield of the $\eta'$ meson, proposed earlier 
as a signal of partial $U_A(1)$ restoration, is shown to create a ``hole'' 
in the low $p_t$ region of $\lambda_*$. 
A comparison with NA44 data from central S+Pb collisions
at 200 AGeV is made and implications for future heavy ion experiments
are discussed. 
\end{abstract}

\narrowtext

{\it \underline{Introduction:}} 
Intensity interferometry is a useful method for studying the
space-time geometry of high energy nucleus-nucleus collisions and 
elementary particle reactions (for recent reviews, see Ref. 
\cite{zajc_93,lorstad_89}). 
In particular, pion interferometry has proved useful in studying the 
space-time dependence of pion emission as was first shown experimentally
by Goldhaber, Goldhaber, Lee and Pais \cite{gglp_60}.
The method of intensity interferometry, known also as 
Hanbury-Brown-Twiss (HBT) correlations, was introduced by Hanbury-Brown 
and Twiss \cite{hbtorig} for measuring the angular diameters of main sequence 
distant stars.   The purpose of this Letter is to show that pion interferometry
can be used to detect the axial $U_A(1)$ restoration and the related 
increase of the $\eta'$ production. 

As was shown in several papers\cite{csorgo_96ch,nickerson_97}, 
at incident beam energies of 200 AGeV 
at the CERN SPS, the space-time structure of pion emission in high 
energy nucleus-nucleus collisions can be separated into two regions: 
the core and the halo.
The pions which are emitted from the core or central region consist 
of two types.  The first type is produced from a direct production 
mechanism such as the hadronization of wounded stringlike 
nucleons in the collision region. 
These pions rescatter as they flow outward with a rescattering time 
on the order of 1 fm/c.   The second type is produced from  
the decays of short-lived hadronic resonances such as the $\rho$, $N^*$,
$\Delta$  and $K^*$, whose decay time is also on the order of 1-2 fm/c.  
This core region is resolvable by Bose-Einstein correlation (BEC) 
measurements.  The halo region, however, 
consists of the decay of long-lived hadronic resonances 
such as the $\omega$, $\eta$, $\eta '$ and $K^0_S$ whose 
lifetime is greater than 20 fm/c.  This halo region is not resolvable by 
BEC measurements. However, as will be summarized below, this region 
still affects the Bose-Einstein correlation function.  

In recent papers \cite{kapusta96etap,xnwang}, 
it was argued that the partial restoration of $U_A(1)$ symmetry of QCD 
and related decrease of the $\eta'$ mass 
\cite{pisarski,kunihiro,shuryak,hatsuda} 
in regions of sufficiently hot and dense 
matter should manifest itself in the increased production of $\eta'$ mesons.  
Estimates of Ref. \cite{kapusta96etap} show that the corresponding 
production cross section of the $\eta'$ should be enhanced by 
a factor of 3 up to 50 relative to that for $p+p$ collisions. 

The effective intercept 
parameter, $\lambda_*$, can be written in terms of the one-particle 
invariant momentum distributions\cite{csorgo_96ch,nickerson_97}
of the core and halo pions and thus is 
sensitive to the abundance of the long-lived hadronic 
resonances such as the $\eta'$.
To see this, consider the two-particle 
Bose-Einstein correlation function which is defined as;
\beq 
C(\Delta k, K) = \frac{N_2({\bf p_1, p_2})}{N_1({\bf p_1}) N_1({\bf p_2})}, 
\eeq{def_becf}
where the inclusive $n$-particle invariant momentum distribution is given as
\beq 
N_n({\bf p_1,...,p_n}) = \frac{1}{\sigma_{in}} E_1...E_n \frac{d\sigma}{
d{\bf p_1}...d{\bf p_n} },
\eeq{lpart}
the relative and the mean four-momenta are given by 
\beq 
\Delta k = p_1 - p_2, \;\;\;\;\;\; K = \frac{p_1 + p_2}{2},
\eeq{def_kK}
and $p = (E_{\bf p},{\bf p})$.

From the four assumptions made in the core-halo model\cite{csorgo_ch97}, 
the Bose-Einstein correlation function is found to be
\beq 
C(\Delta k,K) \simeq  1 + \lambda_*(K = p; Q_{min}) R_c(\Delta k,K),
\eeq{becf2}
where the effective intercept parameter 
and the correlator of the core are defined, respectively, as
\beq 
\lambda_* = \lambda_*(K = p; Q_{min}) = \left [\frac{N_c({\bf p})}
{N_c({\bf p})+N_h({\bf p})}\right ]^2
\eeq{lambda1}
and
\beq 
R_c(\Delta k,K) = \frac{| \tilde{S}_c(\Delta k,K)|^2}{|
\tilde{S}_c(\Delta k = 0,K = p)|^2} \; .
\eeq{correlator}
Here, $\tilde{S}_c(\Delta k,K)$ is the Fourier transform of the 
one-boson emission function, $S_c(x,p)$, 
and the subscripts $c$ and $h$ indicate the contributions from the 
core and the halo, respectively.

In this form, $\lambda_*(K = p, Q_{min})$ is simply related to the momentum
distributions of the core and halo pions.   The $Q_{min}$ 
dependence of $\lambda_*$ which essentially indicates the separation of the 
core and the halo is actually defined by the 
experimental two-track resolution $Q_{min}$.  

{\it \underline{Axial Symmetry Restoration and the $\eta '$ production:}}
In the chiral limit ($m_u = m_d = m_s = 0$), QCD possesses a $U(3)$ chiral 
symmetry. When broken spontaneously, $U(3)$ implies the existence of  
nine massless Goldstone bosons.  In nature, 
however, there are only eight light pseudoscalar mesons, 
a discrepancy which is resolved by the Adler-Bell-Jackiw $U_A(1)$ anomaly; 
the ninth would-be Goldstone boson gets a mass as a consequence of the 
nonzero density of topological charges in the QCD vacuum
\cite{witten,veneziano}.  
In Refs. \cite{kapusta96etap,xnwang}, it is argued
that the ninth (``prodigal'' \cite{kapusta96etap}) Goldstone boson, 
the $\eta'$, would be abundantly produced if sufficiently hot
and dense hadronic matter is formed in nucleus-nucleus collisions.

It was also observed, however, that the $\eta'$ decays are characterized 
by a small signal-to-background ratio in the direct two-photon decay mode. 
This may make the observation of $\eta'$ in this  
mode difficult, especially at small transverse momenta, where 
the increase is predicted to be the strongest. 
However, we now show that the momentum dependence of $\lambda_*$ 
from pion correlations provides a good observable for
partial $U_A(1)$ restoration.

If the $\eta'$ mass is decreased, a large fraction of 
the $\eta'$s will not be able to leave 
the hot and dense region through thermal
fluctuation since they need to compensate for the missing mass by large
momentum \cite{kapusta96etap,xnwang,shuryak}.  
These $\eta'$s will thus be trapped in the hot and dense region until it
disappears, after which their mass becomes normal again and 
as a consequence of this mechanism, 
they will have small $p_t$.  The $\eta'$s then decay to pions via 
\beq
\eta' \rightarrow \eta + \pi^+ + \pi^-  \rightarrow (\pi^0 + \pi^+ + \pi^-)
+ \pi^+ + \pi^-.
\eeq{etap_decay}
Assuming a symmetric decay configuration $(|p_t|_{\pi^+} 
\simeq |p_t|_{\pi^-} \simeq |p_t|_{\eta})$ 
and letting $m_{\eta'} = 958$ MeV, $m_{\eta} = 547$ MeV and 
$m_{\pi^+} = 140$ MeV, the average $p_t$ of the pions from the 
$\eta'$ decay is found to be $p_t \simeq 138$ MeV.
In the core-halo picture the $\eta', \eta$ 
decays contribute to the halo due to their
large decay time ($1/{\Gamma_{\eta', \eta}} > 20$ fm/c).   
Thus, we expect a hole in the low $p_t$ region of the 
effective intercept parameter, 
$\lambda_* = [N_{core}({\bf p})/N_{total}({\bf p})]^2$, centered around
$p_t \simeq 138$ MeV.

We note that the shape of $\lambda_*$ will also 
be effected if the masses 
of the $\omega$ and $\eta$ mesons decrease.
However,  due to the large inelastic cross sections 
for $\omega$ - meson scattering, the $\omega$ 
are expected to rapidly reach chemical equilibrium  
when the hadronic fireball cools and their mass returns to 
its ``normal'' value.  In this case, we do not expect a sizeable
increase in the overall production of the $\omega$ mesons.
In addition, any enhanced production of the $\eta$ mesons should only 
increase the depth of the hole primarily in the $p_t \simeq 117$ MeV region.  
In the case of equal production of the $\eta$ and $\eta'$, there 
will be on the order of twice as many $\pi^+$ coming from the decay of the 
$\eta'$s than from the decay of the $\eta$s.  Thus, we concentrate 
on the dynamics of the $\eta'$ mesons giving an estimated lower bound on 
the depth of the produced hole.

{\it \underline {Description of the Simulation:}}
In the following calculation of $\lambda_*$, we suppress the rapidity
dependence by considering the central rapidity 
region, $( -0.2 < y < 0.2 )$.   
As a function of $m_t = \sqrt{p_t^2 + m^2}$,  $\lambda_*(m_t)$ is given by 
\beq
\lambda_*(m_t) = \left [ \frac{N^{\pi^+}_{core}(m_t)}
{N^{\pi^+}_{total}(m_t)}  \right ]^2,
\eeq{lambda_mt}
where the numerator represents the invariant $m_t$ distribution of $\pi^+$ 
emitted from the core and where the denominator represents the invariant $m_t$ 
distribution of the total number of $\pi^+$ emitted.  
The denominator may be explicitly written as
\begin{eqnarray}
N^{\pi^+}_{total}(m_t) & = &N^{\pi^+}_{core}(m_t) + N_{\omega \rightarrow
\pi^+}(m_t) +  N_{\eta' \rightarrow \pi^+}(m_t) \nonumber\\ 
\null & \null & \,\,\, + N_{\eta \rightarrow
\pi^+}(m_t) +  N_{K^0_S \rightarrow \pi^+}(m_t).
\label{n_total} 
\end{eqnarray}
A detailed analysis \cite{nickerson_97}
has shown that the $\omega$ does not to contribute to the
core in the S+Pb reaction and in the NA44 acceptance.

To calculate the $\pi^+$ contribution from the halo region,  
the bosons ($\omega$, $\eta'$, $\eta$ and $ K^0_S$) are given both 
a rapidity $(-1.0 < y < 1.0)$ and an $m_t$ and then are decayed using
Jetset 7.4 \cite{jetset74_94}.    
The $m_t$ distribution \cite{csorgo_96ch,3d} of the bosons is given by
\beq  
N(m_t) = C m_t^{\alpha} e^{-m_t/T_{eff}}, 
\eeq{mt_dist}
where C is a normalization constant, where $\alpha = 1 - d/2$
and where \cite{3d,na44_teff}
\beq
T_{eff} = T_{fo} + m \langle u_t \rangle^2 . 
\eeq{Teff} 
In the above expression, $d=3$ is the dimension of expansion, 
$T_{fo} = 140$ MeV is the freeze-out temperature 
and $\langle u_t \rangle $ is the average transverse flow velocity. 
The $m_t$ distribution of the core pions is 
also obtained from Eqs. (\ref{mt_dist}) and (\ref{Teff}).  
The contributions from the decay products of the different regions 
(halo and core) 
are then added together according to their respective fractions, allowing for
the determination of $\lambda_*(m_t)$.    
The respective fractions of pions are estimated from both  
the Fritiof \cite{fritiof_87} and the Relativistic Quantum Molecular
Dynamics (RQMD) \cite{rqmd_93} models 
as summarized in Ref. \cite{heiselberg_96}.
The calculation using Fritiof abundances is shown in Fig 2 (solid line).  
A similar $m_t$ dependence but with a slightly
higher value of $\lambda_*(m_t)$ is obtained when using RQMD abundances.

Simulating the presence of the hot and dense region involves including  
an additional relative fraction of $\eta '$ with a medium 
modified $p_t$ spectrum. The $p_t$ spectrum of these $\eta '$ 
is obtained by assuming  
energy conservation and zero longitudinal motion at the boundary between 
the two phases.  This conservation of transverse mass 
at the boundary implies,
\beq
m^{*2}_{\eta'} + {p_t}^{*2}_{\eta'} = m^{2}_{\eta'} + {p_t}^{2}_{\eta'}, 
\eeq{cons_mt}
where the ($*$) denotes the $\eta'$ in the hot dense region. 
The $p_t$ distribution then becomes a twofold distribution.  
The first part of the distribution is from the $\eta'$ which 
have $p_t^* \leq \sqrt{m^{2}_{\eta'} - m^{*2}_{\eta'}}$.   
These particles are given a $p_t = 0$.  
The second part of the distribution comes from the rest of the $\eta'$'s 
which have big enough $p_t$ to leave the hot and dense region.  These have the
same, flow-motivated $p_t$ distribution as the other produced resonances and 
are given a $p_t$ according to the $m_t$
distribution 
\beq 
N_{\eta'}(m^*_t) = C {m^*_t}^{-0.5} e^{-m^*_t/T'},
\eeq{mt*}
where C is a normalization constant and 
where $T' = 200$ MeV and $m^*_{\eta'}$ is the 
effective temperature and mass, respectively, of the hot and dense region.
 
Assuming $m_{\eta'}^* = 500$ MeV in the above scenario, the 
$m_t$ distribution of the $\pi^+$ from the decay of these $\eta'$ ($\eta'
\rightarrow \eta + \pi^+ + \pi^-$) is shown in Fig. 1.   Also shown 
is the $m_t$ distribution of the $\pi^+$ from $\eta'$ 
assuming no hot and dense matter (Eqs. (\ref{mt_dist}) and (\ref{Teff}) 
with $T_{fo} = 140$ MeV, $\langle u_t \rangle = 0.5$ and 
$m_{\eta'} = 958$ MeV).  Comparison of the two distributions
shows the enhancement of the $\pi^+$ in the low $m_t$ region 
which results from the presence of the hot and dense region.

Using three different effective masses for the $\eta'$ in the hot and dense
region, calculations of $\lambda_*(m_t)$ including the hot and dense regions 
are compared to those assuming the standard abundances in Fig 2.   
The effective mass, $m^*_{\eta'} =  738$ MeV, corresponds to an
enhancement of the production cross section of the $\eta'$ by a factor of 3,
while  $m^*_{\eta'} = 403$ MeV and $m^*_{\eta'} = 176$ MeV 
correspond to factors of 16 and 50, respectively.  
The two data points shown are taken from NA44 data
on central $S+Pb$ reactions at the CERN SPS with incident beam energy of 200
AGeV \cite{na44_mt}.  
The lowering of the $\eta'$ mass and the partial chiral restoration result
in a hole in the effective intercept parameter at low $m_t$.
This happens even for a modest enhancement of a factor of 3 
in the $\eta'$ production.
Similar results are obtained when using RQMD abundances.

In addition, $\lambda_*(m_t)$ is calculated using Fritiof abundances
with different average flow velocities in Fig 3.  
Here it is shown that $\lambda_*(m_t)$ can also be a measure of
the average collective flow.  In our calculations, an average flow velocity of 
$\langle u_t \rangle = 0.50$ results in an approximately flat, 
$m_t$-independent shape for
the effective intercept parameter $\lambda_*(m_t)$, if the value
of $\alpha = 1 - d/2 = -1/2$ is kept fixed in Eq. (\ref{mt_dist}). 
Calculations using RQMD abundances result in a similar dependence on 
$\langle u_t \rangle$, but with slightly higher values of $\lambda_*(m_t)$.

A limitation of our study is that we did not include the effects of possible
partial coherence in the $\lambda_*(m_t)$ function.  This is motivated by the
success of completely chaotic Monte Carlo simulations in describing the
measured two-particle correlation functions at the CERN SPS.  However, a recent
study\cite{csorgo_97prl} indicates
that higher order BE symmetrization effects may also result in
a decrease of $\lambda_*(m_t)$ at low $p_t$.  For the present system, this
effect seems to be negligible, about a 1 \% decrease, where the typical
momentum  scale of this effect is $m_t - m = T_{eff}$ 
and where the typical decrease 
is estimated\cite{csorgo_97prl} 
from the measured radius and slope parameters.
The flat shape of our $\lambda_*(m_t)$ distribution results from the 
inclusion of the flow motivated temperature, $T_{eff}$, along with the
effective, $m_t$ dependent volume factor\cite{csorgo_96ch,3d}, $V_* \propto
m_t^{-d/2}$, 
in Eq. (\ref{mt_dist}).   
This flat shape reproduces the published NA44 data and differs from 
earlier theoretical calculations 
where $\lambda_*$ is found only to increase with increasing $m_t$.           

{\it \underline {Summary:}}
Our results reveal an important relationship between partial $U_A(1)$
symmetry restoration and the shape (hole) of the $\lambda_*(m_t)$ 
parameter of the Bose-Einstein correlation function.
We stress that this proposed signal is observed from the transverse 
mass dependence of the strength of the two-particle correlations,  
correlations which are presently being measured 
for fixed target Pb+Pb collisions at the CERN SPS. 
Measurements of two-particle correlations are also being planned for
nuclear collisions at the Relativistic Heavy Ion Collider (RHIC) at BNL as 
well as at the CERN Large Hadron Collider (LHC).

A qualitative analysis of NA44 S+Pb data suggests no visible sign of 
$U_A(1)$ restoration at SPS energies.  In addition, we deduce a 
mean transverse flow of 
$\langle u_t\rangle \approx 0.50$ in S+Pb reactions.
Let us note that the suggested $\lambda_*$-hole signal of
partial $U_A(1)$ restoration cannot be faked in a conventional
thermalized hadron gas scenario, as it is not possible to create
significant fraction of the $\eta$ and $\eta'$
mesons with $p_t\simeq 0$ in such a case.

{\it \underline {Acknowledgments:}}
One of the authors, T.  Cs\"{o}rg\H{o}, would like to express his thanks to
Mikl\'os and Gy\"orgyi Gyulassy for their kind hospitality while at Columbia
University. {\mbox {D. Kharzeev}} is grateful to {\mbox{J. Kapusta}} and 
{\mbox {L. McLerran}} 
for sharing their ideas with him and an enjoyable collaboration, 
and X.-N. Wang for useful discussions.   S.E. Vance would like to thank Urs
Wiedemann and Ulrich Heinz for useful discussions.  

This work was supported by the OTKA Grants T016206,
T024094, T026435, by the NWO - OTKA Grant N25487,
by an Advanced Research Award of the Fulbright Foundation
and by the Director, Office of Energy Research, Division of Nuclear Physics of
the Office of High Energy and Nuclear Physics of the U.S. Department of Energy
under Contract No. DE-FG02-93ER40764.

\begin{figure}[htb]
\psfig{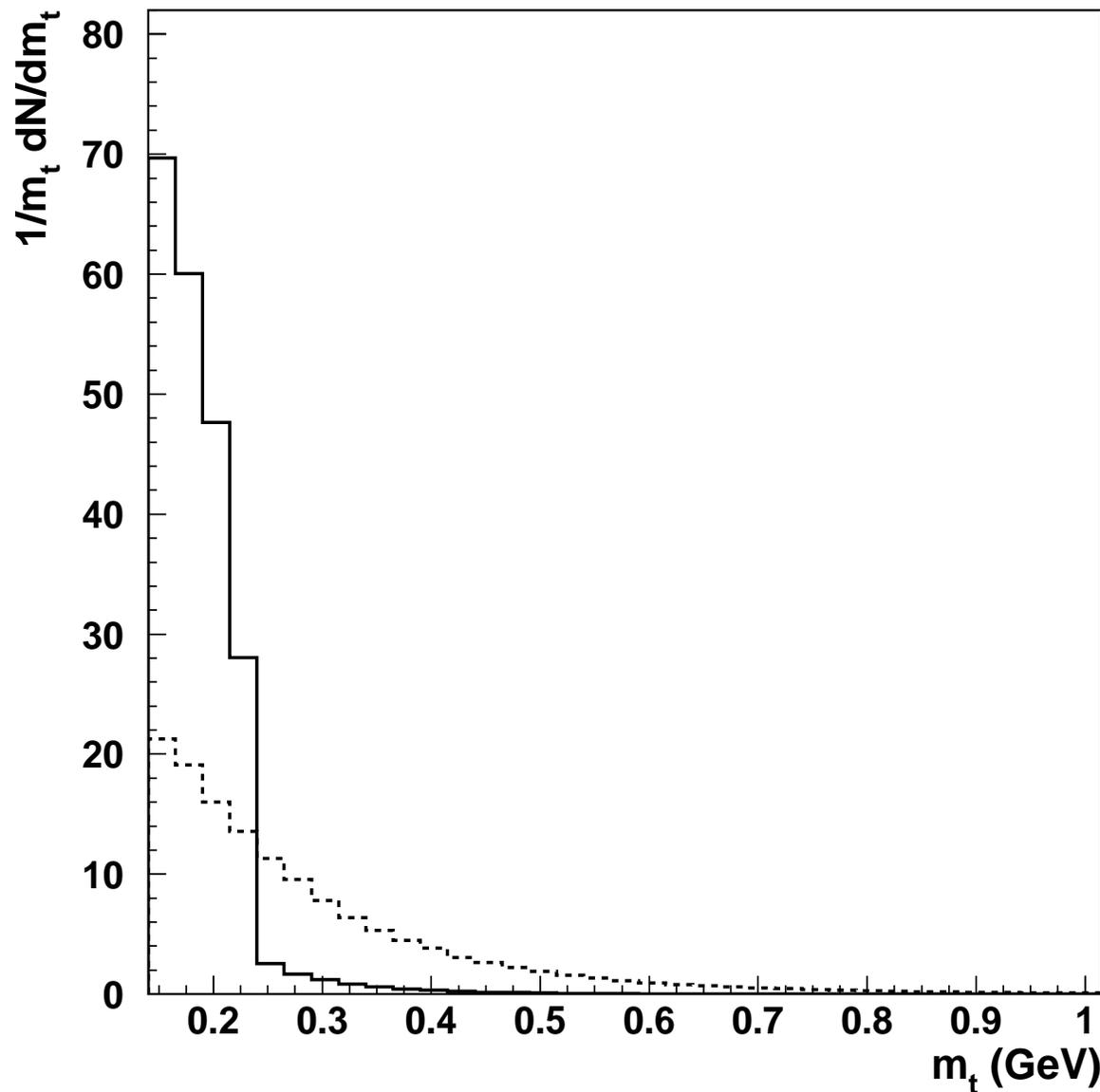}
\caption{Invariant $m_t$ distributions are shown for the
$\pi^+$ from the decay $\eta' \rightarrow \eta + \pi^+ + \pi^-$.  
The solid line assumes the $\eta'$s come from a
hot and dense region where the $\eta'$ have a twofold $p_t$ distribution  
in which $m_{\eta'}^* = 500$ MeV and $T' = 200$ MeV.
The dashed line assumes the $\eta'$ come from nucleus-nucleus collisions at
SPS energies where their $m_t$ spectrum is given by Eqs. (10) and (11), with 
$T_{fo} = 140$ MeV, $\langle u_t \rangle = 0.5$ and $m_{\eta'} = 958$ MeV.} 
\end{figure}

\begin{figure}[htb]
\psfig{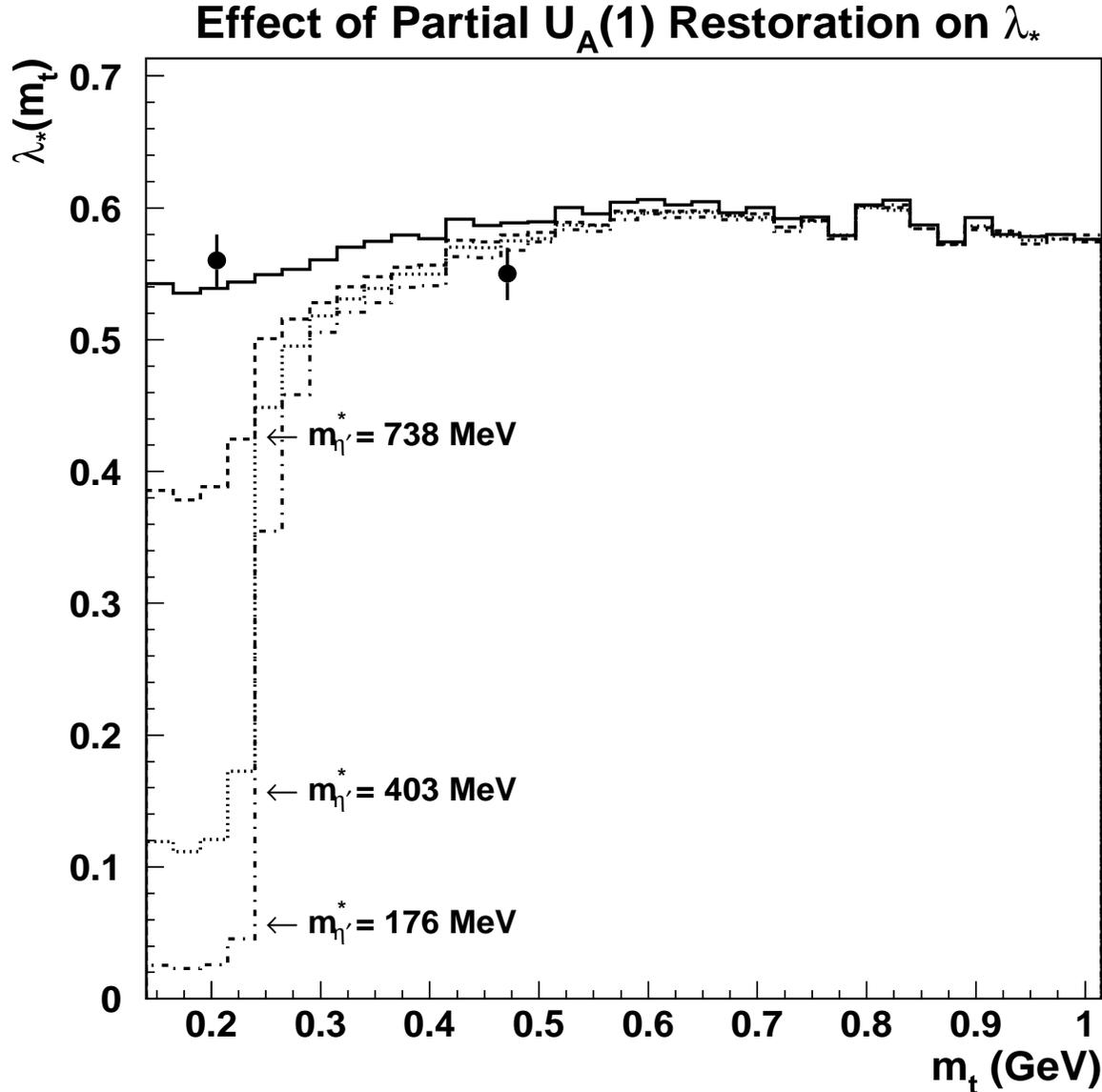}
\caption{Using the estimates of pion abundances given by Fritiof, 
the solid line represents
$\lambda_*(m_t)$ assuming normal $\eta'$ abundances  
while the other lines represent the inclusion of hot and dense regions, 
where $T' = 200$ MeV and the decreased mass of the $\eta'$ is 
$m^*_{\eta'} = 738$ MeV (dashed line), $m^*_{\eta'} = 403$ MeV (dotted line) 
and $m^*_{\eta'} = 176$ MeV (dot-dashed line).  These curves are calculated 
for $\langle u_t \rangle = 0.5$.
The data shown are from S+Pb
reactions at 200 AGeV from the NA44 collaboration.}
\end{figure}

\begin{figure}[htb]
\psfig{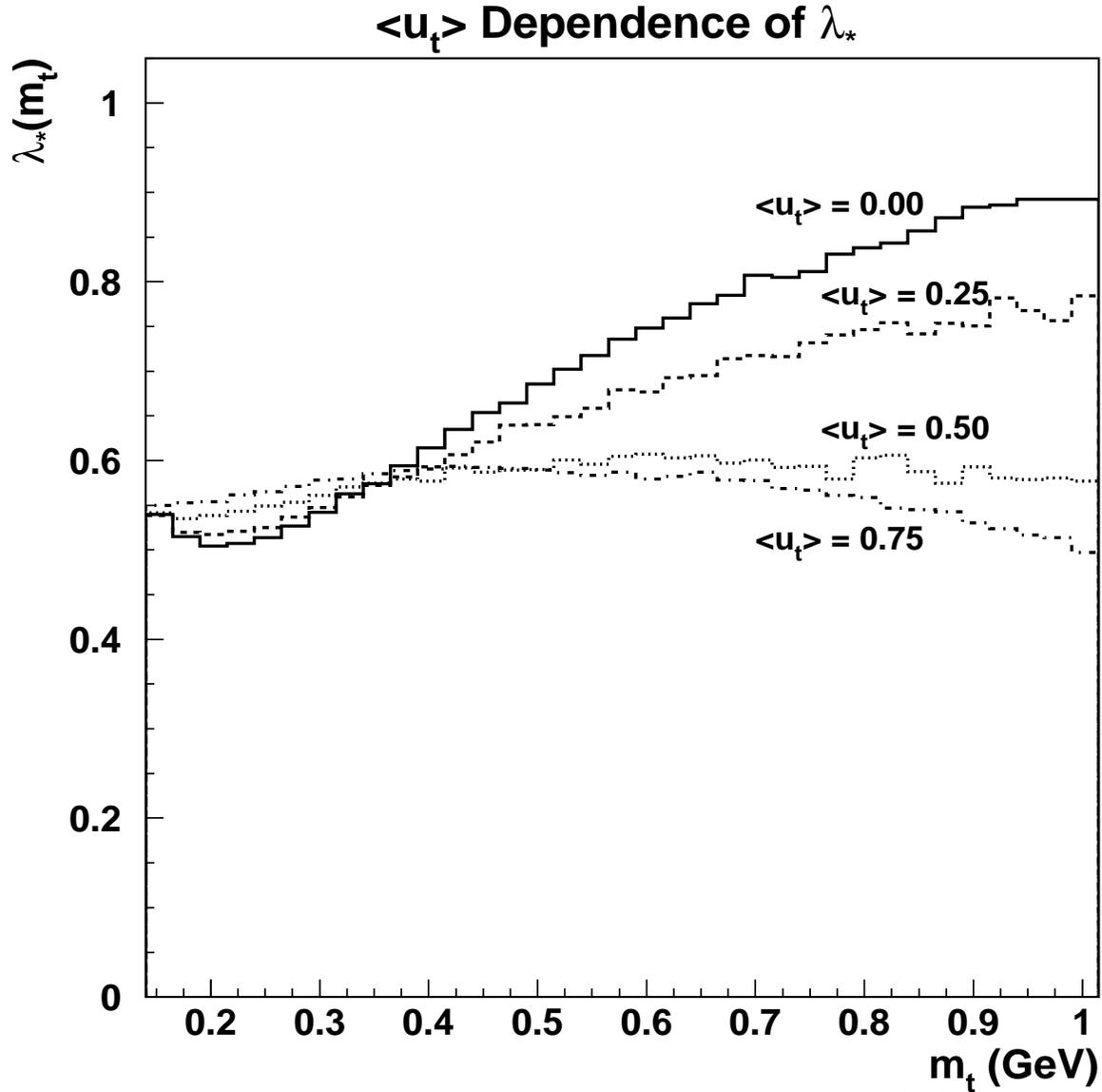}
\caption{Using the estimates of pion abundances given by 
Fritiof, 
 $\lambda_*(m_t)$ is
calculated using different average flow velocities in the $m_t$ distribution.  
It is shown for $\langle u_t\rangle = 0.00$ by the solid line, for
$\langle u_t \rangle = 0.25$ 
by the dashed line, 
for $\langle u_t \rangle = 0.50$ by the dotted line and
for  $\langle u_t\rangle = 0.75$ by the dashed-dotted line.  }
\end{figure}

\end{document}